\documentstyle[aps,preprint,floats,psfig]{revtex}
\tightenlines 

\begin{document}
\draft
\newcommand {\be}{\begin{equation}}
\newcommand {\ee}{\end{equation}}
\newcommand {\bea}{\begin{eqnarray}}
\newcommand {\eea}{\end{eqnarray}}
\newcommand {\nn}{\nonumber}

%
%\twocolumn[\hsize\textwidth\columnwidth\hsize\csname @@twocolumnfalse\endcsname
%
%
% Title Page
%

\title{
Ferromagnetism of $^3$He Films in the Low Field Limit}

\author{H.M. Bozler, Yuan Gu, Jinshan Zhang, K.S. White, C.M. Gould}
\address{Department of Physics and Astronomy, University of Southern
California, Los Angeles, CA 90089-0484}

\date{\today}
\maketitle

\begin{abstract}

We provide evidence for a finite temperature ferromagnetic transition in 2-dimensions
as $H \rightarrow 0$ in thin films of $^3$He on graphite, a model system for
the study of two-dimensional magnetism. We perform pulsed and CW NMR experiments at fields of 
0.03 - 0.48 mT on $^3$He at areal densities of 20.5 - 24.2 atoms/nm$^2$. 
At these densities, the second layer
of $^3$He has a strongly ferromagnetic tendency. With decreasing temperature, 
we find a rapid onset of magnetization that becomes independent of the applied field at
temperatures in the vicinity of 1 mK. Both the
dipolar field and the NMR linewidth grow rapidly
as well, which is consistent with a large (order unity)
polarization of the $^3$He spins.
\end{abstract}
%\vskip2pc]

%\section{}

Two-dimensional magnetic systems order only at zero
temperature when their interactions have continuous symmetry\cite{mermin}. However, the ordering
temperature of these systems may become finite when they have a weak anisotropy
\cite{yafet,friedman2}, 
especially when the weak anisotropy is long ranged. This is the case for two-dimensional
films of $^3$He on graphite\cite{godfrin1,godfrin2}. These films have large exchange rates which can be
described by the multiple spin exchange model over a wide range of coverages\cite{roger1}. Much of 
the interest in these films comes from
the dramatic changes in their structural and magnetic properties caused by small changes in density,
including a change in the sign of the effective exchange\cite{godfrin1,roger2}. 
In addition to exchange energy, these films have nuclear dipolar couplings which are
more than three orders of magnitude smaller than the leading order exchange. Thus $^3$He
films provide a useful test system to observe the behavior of real two-dimensional
magnets in the presence of weak anisotropy.

In our experiments we have evaporated $^3$He
films onto a Grafoil substrate. We report here the results of coverages between 20.5 and
24.2 atoms/nm$^2$. The most striking feature of this range of coverages is the rapidly
increasing ferromagnetic tendency associated with the second layer\cite{godfrin2}. In this
range, the first monolayer is solid with a very low exchange rate\cite{hiroki}, while the second solid
layer is incommensurate with the first monolayer and has a much higher exchange rate. Third layer promotion occurs near 18
 atoms/nm$^2$, and therefore a partial fluid layer lies above the magnetic layer throughout
our coverage region. The Grenoble group has analyzed in detail NMR and heat capacity data
above 4 mK using the multiple spin exchange model (MSE) \cite{roger2,roger3} assuming a 
single incommensurate
phase for the second layer, but the possibility of a mixed solid phase (part incommensurate ferromagnetic, part
low magnetism) has been suggested by other experiments\cite{godfrin2} and has not been excluded in the
detailed analysis of multiple spin exchange\cite{roger2}. In fact, earlier work by Schiffer {\it et al.}\cite{schiffer} 
extending to temperatures far below 1 mK reported results that were more consistent with
a mixed phase.

Previous experiments have demonstrated that a single effective exchange
rate is insufficient to describe both NMR and heat capacity experiments throughout this
region, however as coverages approach 24 atoms/nm$^2$ the nearest neighbor Heisenberg model
works well\cite{godfrin3}. Fitting data to a high temperature spin exchange model with a single
effective exchange constant $J$ remains a convenient (if not precise) method for
characterizing the high temperature data with $J$ ranging
from below 0 (below 20 atoms/nm$^2$) to close to 2 mK at $\sim$ 24 atoms/nm$^2$.

With the exception
of the studies by Friedman {\it et al.}\cite{friedman2,friedman1} on $^3$He which used a SQUID NMR 
technique\cite{friedman3} with
filled pores, all of the previous NMR studies used conventional techniques with applied
fields greater than 6 mT. The Larmor frequencies in those (conventional NMR) experiments
were much smaller than the exchange rates, but nevertheless the applied fields were large
enough to substantially enhance the magnetization in the vicinity of $T \sim J$, effectively
masking any spontaneous magnetization -- if it should occur. On the other hand, the very
low field studies by Friedman {\it et al.} indicated the possibility of spontaneous
magnetization although their results in filled pores are difficult to compare directly
with the numerous experiments using monolayer films.

Our new NMR measurements were
performed in applied fields below 0.5 mT using a SQUID technique similar to the one used
by Friedman {\it et al.} except that an indirect cooling method was employed\cite{yuan}. Grade GTY
Grafoil substrates were diffusion bonded to silver foils for thermal contact. We
lithographically divided the foils into narrow wires prior to the diffusion bonding
process to reduce eddy currents in order to obtain the short pulse recovery times required for 
SQUID NMR and to ensure complete penetration of the time dependent ($H_1$) field. In order to 
minimize eddy current effects, we oriented the static magnetic field perpendicular to the nominal 
plane of the grafoil. In this orientation, the magnetization experiences a competition between 
the static field and the dipolar field and can eventually flop into the grafoil planes at 
sufficiently high polarizations and low magnetic fields. In the data shown in this paper, 
the field and/or temperature is always sufficient to hold the magnetization perpendicular to 
the grafoil sheets. At
higher temperatures we used pulsed NMR, but swept frequency continuous wave
measurements were needed wherever the effective free induction decay time fell below 300 
$\mu$s (typically below $\sim$ 2 mK). Our silver foil contact system allowed us to reach
temperatures as low as 0.3 mK with the lowest temperature being limited by long
equilibrium times.

$^3$He coverages were measured using $^3$He isotherms with an {\it in situ}
pressure gauge located near the sample cell on the nuclear demagnetization cryostat.
Comparing results with other work can be accomplished by anchoring the coverage
scale to the point where the
effective exchange rate crosses zero. This point ranges from approximately 20 atoms/nm$^2$
in references \cite{godfrin2} (Fig. 45) and \cite{schiffer} down to 19.8 atoms/nm$^2$ in 
reference \cite{collin} and 18.9 atoms/nm$^2$ in reference \cite{siqu1}. Our coverages 
give the onset of ferromagnetism at 19.6 atoms/nm$^2$ $\pm$ 2$\%$.

Our experimental results provide evidence for spontaneous magnetization by 1) the
temperature dependence of the magnetization, 2) the rapid increase in the NMR linewidth
at low $T$ with a concomitant shift in the NMR frequency, and 3) the large magnetization observed at
finite temperature even in the low field limit. 

Figure 1 shows the magnetization vs.
temperature for several coverages ranging from densities where the second layer barely shows a
ferromagnetic tendency, to the coverage where the maximum effective exchange has been
observed\cite{godfrin2}. Above 2 mK the magnetization appears quite small. We see that the rapid
increase in magnetization is initiated over a relatively narrow range of temperature. In
addition, the magnetization appears to increase approximately linearly at temperatures
below an ``onset'' temperature near 1 mK. The onset of magnetization remains rounded even
at our lowest applied fields; however, we believe that the rounding is a result of the
considerable heterogeneity of Grafoil samples. The existence of a finite magnetization is
well supported by the frequency shift and linewidth data shown below. 

Another feature of
the magnetization data in Figure 1 is the apparent increase with increasing coverage in
the number of spins participating in the magnetization -- even at our lowest temperatures.
Since the second layer density only increases by a few percent \cite{godfrin1}, we 
interpret this increasing magnetization as evidence that for coverages below 24 atoms/nm$^2$, we have a mixed phase
as suggested by Schiffer {\it et al.} \cite{schiffer} 

Below the onset temperature, we observe a rapid increase
in the NMR linewidth shown in Figure 2. This increase is readily observed as a decrease in the effective $T_2$
of the free induction decay signals from pulsed measurements, and then in direct
measurements of the linewidth in the CW measurements at lower temperatures. Typical NMR
lines are shown in the inset in Figure 2. Along with the increasing amplitude and linewidth
we see a frequency shift caused by the increasing dipolar field from the aligned $^3$He spins 
and the appearance of a characteristic distortion in the lineshape at temperatures slightly
below 1 mK with the central line shifting towards lower frequencies while portions
of the NMR line extend to frequencies above the Larmor frequency. These extremely broad NMR 
lines were also observed in reference \cite{schiffer}. 

In the simplest case where the static field and polarization of two-dimensional $^3$He spins are perpendicular to 
the graphite planes, a dipolar field ($\lambda M_0$) results in a negative frequency shift. Much of the 
characteristic lineshape that we observe at low temperatures can be attributed to the distribution of 
plate angles for surfaces in Grafoil. These angular distributions have been studied by neutron scattering
\cite{godfrin1,schildberg,kjents} and x-ray scattering \cite{bouldin} with the probability distribution 
 shown to be a gaussian centered around the normal ($\theta=0$) to the Grafoil plus a 
 randomly distributed portion so that 
$P(\theta)/sin(\theta)=a\exp(-\theta/0.3145)^2+b$ where $a$ and $b$ are normalized to fraction
of spins in the central distribution and at random angles. Reported values of the random fraction range from 50$\%$ 
\cite{godfrin1,schildberg} down to 33$\%$ \cite{bouldin}. In order to calculate the distribution
of frequencies that simulate our NMR lines we need to calculate the angle of the magnetization 
(which is not exactly the angle of the applied field at low fields) and then the resulting NMR
frequency assuming 1) a temperature dependent average dipolar field ($\lambda M_0$), 2) the equations for spin dynamics 
from Friedman {\it et al.} \cite{friedman1}, and 3) a temperature dependent intrinsic linewidth. We can approximately fit our 
NMR absorption lines to this model, however this process works best when we assume that only
20-25$\%$ of the spins are on the randomly distributed plates. We find that fits to the data work best
when the intrinsic lineshape
is described by a Gaussian rather than a Lorentzian (especially at temperatures below 1 mK). 
This fit results in estimates for the 
Gaussian linewidth and dipolar field produced by the $^3$He spins.

Figure 2 shows the effective linewidth resulting from fits to the NMR lines. We can see
that the NMR linewidth grows rapidly below the onset temperature. We show data at 22.5 atoms/nm$^2$ 
but the linewidths at other coverages are very similar in both magnitude and 
temperature dependence. A further contribution to the linewidth (and shape) is the variation of
the local dipolar field across a finite sheet and at edges.

The negative shift of the maximum in the absorption peak is a measure of the dipolar field. In 
 Figure 3 we plot the dipolar field $\lambda M_0$ based on our fits to the NMR lines. 
 We see that the dipolar field increases rapidly
below the onset temperature. We also see that the dipolar field as well as the linewidth are proportional to
the magnetization in Figure 1 which was obtained by directly integrating the NMR lines. The
magnitude of the dipolar field extrapolates
to close to 1.5 mT at zero temperature which is somewhat less than the dipolar field that we 
would expect for a fully polarized sheet of spins at the second monolayer density but comparable to
the shifts observed by Schiffer {\it et al.}\cite{schiffer} at similar coverages with much higher
applied fields. The dipolar field from the higher (24.2 atoms/nm$^2$) coverage is approximately 10$\%$ 
greater. 

In Figure 4 we show the measured magnetization over a wide range of temperatures and at
four magnetic fields on a logarithmic scale. An important aspect of the data in Figure 4
is the enormous (more than four orders of magnitude) increase in magnetization. In fact,
if we normalize our observed signals to the calculated high temperature polarization (including
contributions from the first and fluid layers), we
obtain low temperature polarizations of the second layer which are very close to 100$\%$. 
At the lowest applied field (0.03 mT) there is a very
high slope near 1.3 mK (magnetization doubling each 10$\%$ change in temperature) and near that 
temperature we see a
crossover from the high temperature region where the magnetization is proportional to the
magnetic field, to the low temperature region where the magnetization is independent of
the field. 

The question arises whether the ordering which we observe is intrinsic to the
$^3$He/Grafoil system and thus a property of such 2-dimensional magnetic systems, or whether
it is caused by some artifact of an imperfect substrate. We note that at coverages below
19.5 atoms/nm$^2$, we did not observe any evidence of a low temperature order. Thus, the
ordered system in the low field limit that we observe coincides in coverage with 
the change in sign of the effective exchange.
If the order were solely due to defects in the substrate, it would seem unlikely that
these effects would disappear within this tiny change in coverage. In principle, heat
capacity measurements should also indicate the existence of a magnetic ordering. In fact,
measurements in this coverage range have been used to argue that no such order exists\cite{ishida}.
However, it is likely that any divergence in the heat capacity would be extremely weak
making it extremely difficult to observe anomalies when samples are heterogeneous. 

The dominant interaction between $^3$He atoms in these films is particle exchange which is well
described by the MSE model\cite{roger1}. By itself, this model predicts phase
ordering only at zero temperature, consistent with the Mermin-Wagner theorem\cite{mermin,kop}. However 
weak anisotropies modify the spin-wave spectrum, so that long wavelength magnons
no longer destroy order \cite{yafet,mills}. 
Recently there has been a suggestion of a large anisotropic effect by Gov {\it et al.}
\cite{gov} due to correlations in zero-point motion, however this proposal requires a substantial
modification of the MSE model and we do not yet have a way to experimentally test this
idea.

Yafet {\it et al.} \cite{yafet} calculated the reduction of magnetization caused by spin waves in the presence of 
dipole-dipole interactions. They showed that these interactions introduce a linear term in 
the spin wave spectrum that in turn stabilizes the ordered state. Their numerical estimates for the
magnetization (in a square lattice) have an approximately linear temperature dependence consistent
 with our data.

The reduction of magnetization by spin waves in the pure Heisenberg model has been calculated by Kopietz 
{\it et al.}\cite{kop}.
In the case of metal films with weak anisotropy this magnetization is described by
a different relation\cite{mills} 
$\delta M/M_0 \equiv 1-M/M_0 =1-(cT/4\pi J) \ln (T/\delta H)$ where $\delta H$ is the applied field minus the
dipolar field in appropriate units and $c$ is a constant of order unity. Because of the competition between exchange cycles in the MSE model, 
the correct value of $J$ to use is uncertain. Nevertheless putting in our values for $J \simeq $ 2.2 mK 
(at our highest coverage) and $\delta H$ corresponding to our applied field minus our measured dipole field, we 
get $\delta M/M_0$ to fit our data when $c=0.5$. Clearly a more complete theory which
includes calculations on a triangular lattice is needed to complete a quantitative comparison.
In addition, experiments on a more homogeneous substrate could clarify the nature of the ferromagnetic 
ordered state.

We wish to acknowledge the assistance of Barry Fink and useful discussions with Douglas Mills. This
research is supported by the National Science Foundation grant DMR-9973255.

\begin{figure}[h]
\centerline{\psfig{figure=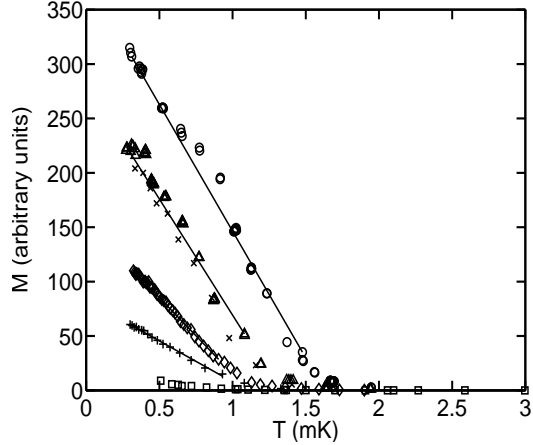,width=7cm,height=6.0cm,angle=0}}
\vspace{0.5cm}
\caption{
Magnetization as a function of temperature for a variety of coverages.
($\circ$) 24.2, (x, $\triangle$) 22.5, ($\Diamond$) 21.8, (+) 21.3, ($\Box$) $20.5$ atoms/nm$^2$.
All data were taken at 0.35 mT, except for the lowest coverage data which was taken at 0.48 mT. 
}
\end{figure}

\begin{figure}
\centerline{\psfig{figure=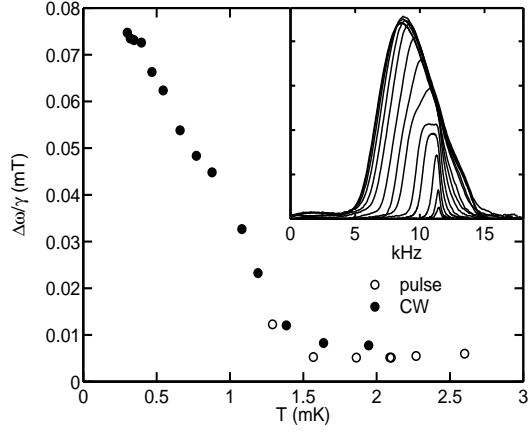,width=7cm,height=6.0cm,angle=0}}
\vspace{0.5cm}
\caption{
Linewidth as a function of temperature for a coverage of 22.5 atoms/nm$^2$ at 0.35 mT based on the fitting
procedure described in the text.
$\gamma$ is the gyromagnetic ratio for $^3$He.
The inset shows typical absorbtion lines taken from CW NMR ranging from 2 mK
(smallest) down to 0.3mK (largest).
}
\end{figure}

\begin{figure}
\centerline{\psfig{figure=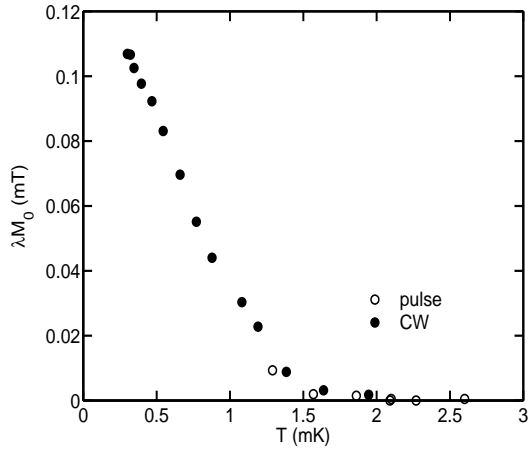,width=7cm,height=6.0cm,angle=0}}
\vspace{0.5cm}
\caption{
Average dipole field (NMR resonance shift) $\lambda M_0$ as a function of temperature for a 
coverage of 22.5 atoms/nm$^2$ at 0.35 mT.
}
\end{figure}

\begin{figure}
\centerline{\psfig{figure=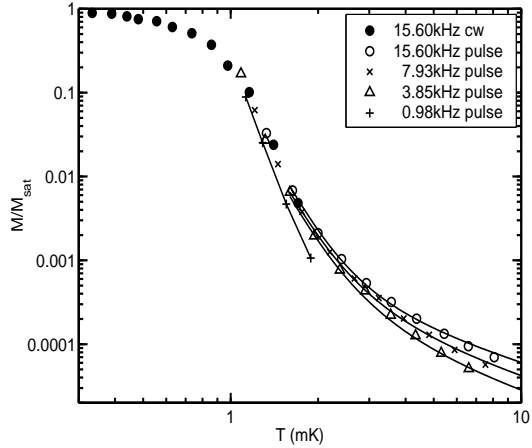,width=7cm,height=6.0cm,angle=0}}
\vspace{0.5cm}
\caption{
Total magnetization at 22.5 atoms/nm$^2$. The open symbols were from pulsed NMR 
with applied fields of ($\circ$) 0.48 mT, (x) 0.24 mT, ($\triangle$) 0.12 mT, (+) 0.03 mT.
The closed symbols ($\bullet$) were from CW NMR at 0.48 mT. Additional CW data (not shown) 
from lower fields agrees well with the 0.48 mT results. $M_{sat}$ was the estimated low temperature
limit of $M$. The solid lines while intended as ``guides to the eye'' were the result of a high temperature
 Heisenberg Model expansion in the case of the three largest fields. 
}
\end{figure}

\end{document}